\documentstyle[11pt,newpasp,twoside,epsf]{article}
\def\edcomment#1{\iffalse\marginpar{\raggedright\sl#1\/}\else\relax\fi}
\reversemarginpar

\def\beq{\begin{equation}}
\def\eeq{\end{equation}}
\def\ber{\begin{eqnarray}}
\def\eer{\end{eqnarray}}
\def \o {\Omega}

\def\omegam{\Omega_{\rm M}}

\def\omegal{\Omega_{\phi}}
\def\omega0{\Omega_0}
\def \lleq {\lower0.9ex\hbox{ $\buildrel < \over \sim$} ~}
\def \ggeq {\lower0.9ex\hbox{ $\buildrel > \over \sim$} ~}
\def \l {\Lambda}
\def \ap {ApJ, }

\def \pd {Phys. Rev. D, }

\def\simle{\mathrel{\spose{\lower 3pt\hbox{$\mathchar"218$}}
     \raise 2.0pt\hbox{$\mathchar"13C$}}}
\def\simge{\mathrel{\spose{\lower 3pt\hbox{$\mathchar"218$}}
	  \raise 2.0pt\hbox{$\mathchar"13E$}}}

\begin{document}

\title {Scalar field models for an accelerating universe}
\author{Varun Sahni}
\affil{Inter-University Centre for Astronomy \& Astrophysics,
Post Bag 4, Pune 411007, India}

\begin {abstract}
I describe a new class of quintessence+CDM models in which late time
scalar field oscillations can give rise to both quintessence
and cold dark matter. Additionally, a versatile ansatz 
for the luminosity distance is used to reconstruct the quintessence
equation of state in a {\em model independent} manner from observations of high
redshift supernovae.
\end {abstract}

\section{A new model of quintessence and cold dark matter}
The supernova-based discovery that the universe may be accelerating 
can be explained within a general relativistic framework provided
one speculates the presence of a matter component with negative pressure,
the most famous example of which is the cosmological constant `$\l$'
(Perlmutter et al. 1998,1999; Riess et al. 1999).
$\Lambda$ runs into formidable fine tuning problems since its value
must be set $\sim 10^{123}$ times smaller than the
energy density in the universe at the Planck time in order to
ensure that $\l$ dominates the total energy density at
precisely the present cosmological epoch. This involves a fine tuning 
of one part in $10^{123}$ at the Planck scale or one part in $10^{53}$ at the 
Electroweak scale. 

One way around this difficulty is
to make $\l$ time-dependent, perhaps by using scalar field models which 
successfully generate a time-dependent $\l$-term during an early Inflationary
epoch. 
In this context the exponential potential provides an interesting illustration,
since the density in the $\phi$-field tracks
the background matter/radiation density when the latter is cosmologically
dominant (Ratra \& Peebles 1988, Wetterich 1988, Ferreira \& Joyce 1997):
\beq
\frac{\rho_\phi}{\rho_{B} + \rho_\phi} = \frac{3(1 + w_B)}{p^2\lambda^2} \ll 1
\label{eq:ratra}
\eeq
($w_B = 0, ~1/3$ respectively for dust, radiation).
This behaviour allows $\rho_\phi$ to be 
fairly large initially.
Based on this property we introduce 
a new class of cosmological models which can describe both a time-dependent
$\l$-term (quintessence) and cold dark matter (CDM) within the unified framework
of the class of potentials (Sahni \& Wang 2000)
\beq
V(\phi) = V_0(\cosh{\lambda\phi} - 1)^p.
\label{eq:pot1}
\eeq
$V(\phi)$ has asymptotic forms:
\ber
&V(\phi) ~&\simeq ~\tilde{V}_0e^{-p\lambda \phi} ~~{\rm for} ~
 \vert\lambda\phi\vert \gg 1
~(\phi < 0),
\label{eq:exp1}\\
&V(\phi) ~&\simeq ~\tilde{V}_0 (\lambda\phi)^{2p} ~~{\rm for}
~\vert\lambda\phi\vert \ll 1
\label{eq:exp2}
\eer
where $\tilde{V}_0 = V_0/2^p$. The exponential form of $V(\phi)$ guarantees
that the scalar field equation of state mimics background
matter at early times so that $w_\phi \simeq w_B$. 
At late times oscillations of 
$\phi$ lead to a mean equation of state
\beq
\langle w_\phi\rangle = \bigg\langle \frac{\frac{1}{2}{\dot\phi^2} - V(\phi)}
{\frac{1}{2}{\dot\phi^2} + V(\phi)} \bigg\rangle = \frac{p - 1}{p + 1},
\label{eq:w}
\eeq
resulting in cold dark matter with $\langle w_\phi\rangle \simeq 0$ if $p = 1$,
or in quintessence with $\langle w_\phi\rangle \leq -1/3$
if $p \leq 1/2$. We therefore have before us the attractive possibility of 
describing CDM and quintessence in a common
framework by the potential 
\beq
V(\phi,\psi) = V_{\phi}(\cosh{\lambda_{\phi}\phi} - 1)^{p_{\phi}}
   + V_{\psi}(\cosh{\lambda_{\psi}\psi} - 1)^{p_{\psi}}
\label{eq:potCDM}
\eeq
where $p_{\psi} = 1$ in the case of CDM
and $p_{\phi} \leq 0.5$ in the case of quintessence.
In figure 1 we show a working example of this model which agrees well with
observations of high redshift supernovae and does not suffer from
the fine tuning problem faced by $\l$, since $\rho_\phi$ can be fairly large
initially. 
We should add that most models of quintessence
usually work under the assumption that the three matter fields:
baryons, CDM \& quintessence need not be related in any fundamental way and 
might even have different physical origins.
If this is indeed the case then it remains somewhat of a mystery as to
why $\o_\phi$, $\o_m$, (and possibly $\o_b$) have comparable magnitudes at the
present time. By combining quintessence and CDM within
a single class of potentials we make a small step in answering this question
by showing that 
unified models of quintessence and CDM 
are conceivable (Sahni \& Wang 2000).

\begin{figure}
\plotfiddle{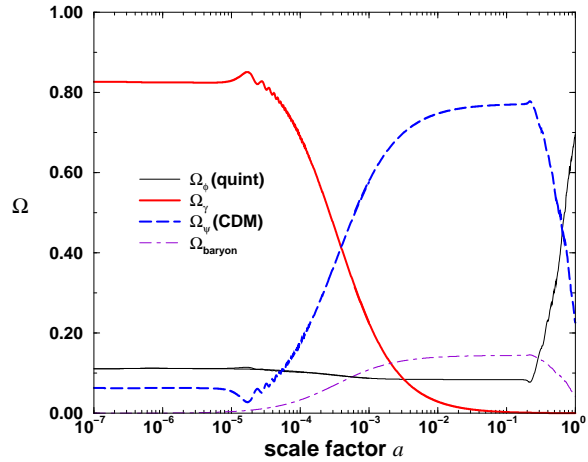}{8cm}{0}{45}{45}{-130}{0}
\caption{\small{The evolution of the dimensionless density parameter for
the CDM field $\Omega_{\psi}$ (dashed line) and quintessence field
$\Omega_{\phi}$ (thin solid line).  Baryon (dash-dotted line) and radiation
densities (thick solid line) are also shown.  For more details see
Sahni and Wang (2000).}}
\end{figure}

An intriguing property of cold dark matter based on (6) is that
it can have a large Jeans length which leads to suppression (frustration)
of clustering on kiloparsec scales. {\em Frustrated Cold Dark Matter} (FCDM)
redresses certain shortcomings of the standard CDM scenario and might
provide a natural explanation for the dearth of dwarf galaxies seen in our
local neighborhood (Sahni \& Wang 2000). 

Other quintessence potentials include
$V(\phi) \propto \phi^{-\alpha}$ (Ratra \& Peebles 1988),
$V(\phi) \propto e^{\beta\phi^2}\phi^{-\alpha}$ (Brax \& Martin 2000)
and $V(\phi) \propto \sinh ^{-2p}(\phi+\phi_0)$ (Sahni \& Starobinsky 2000). 
The latter 
describes quintessence which maintains a constant equation of state
$w = -(1+p)^{-1}$ throughout the matter dominated epoch and later, during
acceleration.

\section{Reconstructing quintessence from
supernova observations}

Although a large class of scalar potentials can describe 
a time dependent
$\l$-term, 
no unique potential has so far emerged from a consideration of
high energy physics 
theories such as supergravity or M-theory.
(The situation in many respects resembles that faced by the 
Inflationary scenario,
for a review see Sahni \& Starobinsky 2000.)
It is therefore meaningful to try and reconstruct $V(\phi)$ directly
from observations in a model independent manner. This is easy to do 
if one notes that,
in a flat FRW universe, the luminosity distance determines
the Hubble parameter uniquely (Starobinsky 1998, Saini et al. 2000)
\begin{equation}
H(z)\equiv \frac{\dot{a}}{a}
=\left[ \frac{d}{dz} \left( \frac{D_L(z)}{1+z} \right) \right]^{-1}.
\label{eqn:hfz}
\end{equation}
The Einstein equations can be written in the suggestive form
{\setlength\arraycolsep{2pt}
\begin{eqnarray}
{8\pi G\over 3H_0^2} V(x)\ &=& {H^2\over H_0^2}
-{x\over 6H_0^2}{dH^2\over dx} -{1\over 2}\omegam\,x^3,
\label{eqn:Vzed}\\
{8\pi G\over 3H_0^2}\left({d\phi\over dx}\right)^2 &=&
       {2\over 3H_0^2 x}{d\ln H\over dx}
  -{\omegam x\over H^2},\label{eqn:phidot}
\end{eqnarray}}
where $x\equiv 1+z$.
Thus knowing $D_L$ we can determine  both $H(z)$ and $dH(z)/dz$, and hence 
$V(\phi)$. The cosmic equation of state can also be 
reconstructed from $D_L$ since
\begin{eqnarray}
w_\phi (x) \equiv {p\over\rho} &=&
\frac{(2x/3) d\ln H/dx -1}{1-\left(H^2_0/H^2\right)
\omegam x^3}.\label{eqn:wzed}
\end{eqnarray}

In order to apply our method to observations we use the following
rational ansatz for the luminosity distance
\begin{equation}
{D_L\over x}
\equiv \frac{2}{H_0}\left[ \frac{x - \alpha\sqrt{x} -1 + \alpha}{\beta x+
\gamma\sqrt{x} + 2 - \alpha -\beta -\gamma}\right]
\label{eq:star}
\end{equation}
where $\alpha$, $\beta$ and $\gamma$
are fitting parameters. This function reproduces the {\em exact} analytical
form of $D_L$ when $\omegal=0, \omegam=1$ and when $\omegal=1, \omegam=0$.
It also has the correct asymptotic behaviour
$H(z)/H_0 \to 1$ for $z\to 0$,
and  $H(z)/H_0 \to (1+z)^{3/2}$ for $z\gg 1$.
Applying a maximum likelihood technique to $D_L$ given by (11) and 
$D_L^{obs}$ obtained from observations
of high redshift supernovae, we can reconstruct $H(z)$, $V(\phi)$ and
$w_\phi(z)$. Our results for $w_\phi$ shown in fig. 2 
indicate some evidence of possible evolution in $w_\phi$
with $-1 \leq w_\phi \lleq -0.80$ preferred at the present epoch, and 
$-1 \leq w_\phi \lleq -0.46$ at $z=0.83$, the farthest SN in the
sample (both at 90\% CL). 
However, a cosmological {\it constant} with $w=-1$ is
also consistent with the data.


\begin{figure}
\plotfiddle{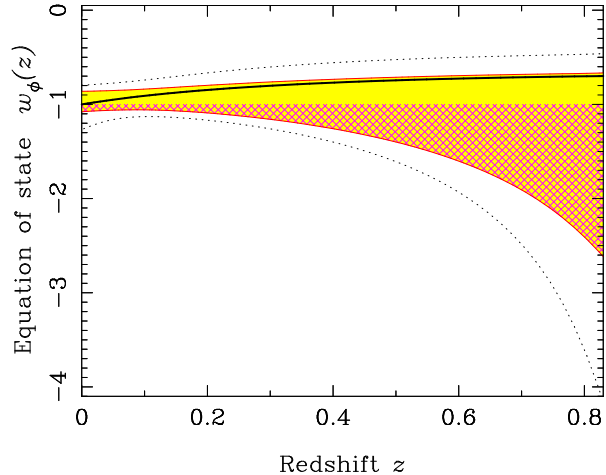}{6cm}{-90}{35}{35}{-130}{200}
\caption{\small {The equation of state parameter $w_\phi(z)= p_\phi/\rho_\phi$ 
as a function of redshift.  The solid line corresponds to the best-fit
values of the parameters.  The shaded area covers the range of 68\%
errors, and the dotted lines the range of 90\% errors.
The hatched area
represents the region $w_\phi \leq -1$, which is disallowed for a
minimally coupled scalar field (from Saini et al. 2000).}}
\end{figure}

\acknowledgements 

The results presented in this talk were obtained in collaboration with 
Somak Raychaudhury, Tarun Saini, Alexei Starobinsky and Limin Wang
whom I would like to thank for many enjoyable discussions.

\end {document}